# Resonant tunneling spectroscopy of valley eigenstates on a hybrid double quantum dot

T. Kobayashi[1, a)], J. van der Heijden[1], M. G. House[1], S. J. Hile[1], P. Asshoff[1], M. F. Gonzalez-Zalba[2], M. Vinet[3], M. Y. Simmons[1], and S. Rogge[1]

[1]*Centre for Quantum Computation and Communication Technology, University of New South Wales, Sydney 2052 New South Wales, Australia*

[2]*Hitachi Cambridge Laboratory, J. J. Thomson Avenue, Cambridge CB3 0HE, United Kingdom*

[3]*Université Grenoble-Alpes and CEA, LETI, MINATEC, 38000 Grenoble, France*

We report electronic transport measurements through a silicon hybrid double quantum dot consisting of a donor and a quantum dot. Transport spectra show resonant tunneling peaks involving different valley states, which illustrate the valley splitting in a quantum dot on a $Si/SiO_2$ interface. The detailed gate bias dependence of double dot transport allows a first direct observation of the valley splitting in the quantum dot, which is controllable between 160-240 μeV with an electric field dependence $1.2 \pm 0.2$ meV/(MV/m). A large valley splitting is an essential requirement to implement a physical electron spin qubit in a silicon quantum dot.

Electrons confined in silicon nanostructures satisfy some crucial requirements for the physical implementation of quantum computation such as their long spin coherence time and excellent controllability [1-4]. To isolate two electronic states used as qubit bases from other states, the degeneracy of the silicon conduction band minima (valleys) must be lifted. Recent investigations have revealed that a large spin-valley hybridization relaxes electron spin qubit states rapidly [5, 6], increasing the importance of controlling the valley splitting [5, 7-11]. While electric field controllability of the valley splitting in a quantum dot (QD) has been evaluated by measuring spin relaxation [5], the direct measurement by transport spectroscopy in a QD has not been reported.

In this paper, we report transport measurements through a silicon hybrid double QD (DQD) consisting of a phosphorus donor and an electrostatically defined QD [12]. The donor ground level is electrically tuned through resonance with the QD valley ground and excited states, which results in two separated resonant tunneling peaks. From the gate voltage dependence of the separation between these peaks, we evaluate the electric field dependence of the valley splitting in the QD. The obtained electric field dependence of 1.2 meV/(MV/m) is one order of magnitude larger than a previous report on a thermal $Si/SiO_2$ interface [5], being comparable with that on a $Si/SiO_2$ interface defined by oxygen implantations [7]. Such a large electric field dependence provides an effective experimental means to tune the valley splitting in situ.

---

a) t.kobayashi@unsw.edu.au

The valley degree of freedom plays an important role in the electronic states in QDs formed in a silicon-on-insulator (SOI) structure. For a free electron in bulk silicon, all six valleys labelled by $\pm k_x$, $\pm k_y$ and $\pm k_z$ are energetically degenerate. Here $x$, $y$ and $z$ axes are chosen to be along the crystallographic directions [100], [010] and [001], respectively. In a two-dimensional electron system confined to the (001) plane by a SOI interface, the anisotropy in the effective mass lifts the sixfold degeneracy: the resultant states are the twofold-degenerate lower energy state ($|\pm k_z\rangle$) and the fourfold-degenerate higher energy state ($|\pm k_x\rangle$ and $|\pm k_y\rangle$) with an energy gap of several tens of meV [13]. Furthermore, the sharp confinement potential of the SOI interface mixes $|\pm k_z\rangle$ states into the valley ground and excited states, $|g\rangle$ and $|e\rangle$ [14-16]. These low-lying eigenstates are energetically split by an energy gap of the order of 0.1 meV on the thermally defined Si/SiO$_2$ interface referred to as valley splitting [5, 6, 8]. An electric field perpendicular to the SOI interface changes the coupling of $|\pm k_z\rangle$ states, changing the size of the valley splitting [5, 7-11]. In a QD defined on the SOI interface, each of its orbital levels with typical level spacings of 0.1-10 meV is further split by this valley splitting [6, 17]. On a donor site, on the other hand, the valley ground state $|A\rangle$ [18] and excited states are gapped by excitation energies larger than 10 meV [19]. This gap is of the same order as the donor orbital excitation energy, which is larger than 30 meV [20]. While it is reported that an electric field perpendicular to the SOI interface can induce a transition of the electronic wave function from donor-like to QD-like, such an effect is only caused by an electric field larger than the Coulomb field [17, 21, 22], which is not the case in this experiment.

Inter-dot resonant tunneling in a DQD occurs when an energy level in one site coincides with that in another site, widely used to measure DQD energy level spectra [23, 24]. In typical transport measurements with a source-drain bias of a few meV, there are no donor excited states accessible, in contrast with a QD which has several accessible states including the $|g\rangle$ and $|e\rangle$ valley states and orbital states. As a consequence, we can employ the energetically isolated donor $|A\rangle$ state as a probe of the QD orbital and valley eigenspectrum.

The device used in this study is a silicon field effect transistor consisting of a silicon nano-wire etched from a 20-nm-thick (100) silicon film with $10^{17}$ cm$^{-3}$ phosphorus concentration on a SOI wafer. The SOI structure is fabricated by wafer bonding with hydrogen implantations (the Smart-Cut® technology) [25]. Characteristics of devices fabricated from the same SOI wafer have already been investigated in previous works [22, 26]. The front gate stack fabricated on the silicon film defines a channel region of 60 nm × 60 nm, which contains ~7 phosphorus dopants. The electronic energy levels of dopant and QD sites in the channel are tuned by front and back gate biases, $V_{FG}$ and $V_{BG}$. The source and drain regions are heavily n-doped. Transport through the channel is measured by applying a bias voltage $V_S$ to the source lead, while the drain lead is kept grounded [Fig. 1(a)]. To perform radio-frequency (rf) reflectometry measurements simultaneously, we attach an inductor of 1.2 μH to the source lead [26, 27]. This forms a tank circuit with a parasitic capacitance of ~0.5 pF and a resonant frequency of 205 MHz, while we apply a slightly detuned rf signal (216 MHz) to the tank as a carrier signal. The carrier power is kept small enough to avoid any change in the transport measurements. The reflectometry probes any change in the rf reflection coefficient around the resonant frequency, which is caused by electron tunneling between a QD or donor site and a lead, driven by the rf power. The rf signal applied to the source lead is partially screened by the wrap-around front gate. This suppresses the reflectometry signal from sites that are only coupled to the drain lead. By combining this information with the transport signal, which is only observed for sites tunnel-coupled to both the source and drain leads, this technique uncovers the location of the individual sites between which electronic transitions occur. All measurements were carried out in a dilution refrigerator with base temperature of ~50 mK.

Figures 1(b) and (c) show typical transport and reflectometry spectra, respectively, over a wide range of $V_{FG}$ and $V_{BG}$ with $V_S = -1$ mV. At large $V_{FG}$ ($V_{BG}$), both transport and reflectometry signals simultaneously show almost periodic peak structures denoted by solid (dashed) lines, which indicates QD sites on the SOI interface of the front (back) gate side. Apart from these periodic structures, the reflectometry signal exhibits two peaks along dotted lines labelled $\alpha$ and $\beta$. These features have different slopes from each other and the QDs, therefore we attribute them to two different donor sites [26].

Figure 2(a) shows a detailed transport spectrum with $V_S = -1$ mV around the region enclosed by the white circle in Fig. 1(b). In this region, we observe a bias triangle structure, indicating series transport through two tunnel-coupled sites. One site is identified as the donor site $\alpha$ by correspondence between the slope of the side of the triangle indicated by a dotted line and that of the donor site $\alpha$ in Fig. 1(c). Single-site transport through another site is manifested as a weak transport feature along the dashed line. From the similarity of the slope of this line to the dashed lines in Figs. 1(b) and (c), we identify this site as a back gate side QD. To determine the detailed arrangement of the donor and QD sites, we focused on the difference between transport and rf reflectometry measurements. The single site transport through the donor $\alpha$, which is manifested as the transport signal along the dotted line $\alpha$, is much weaker than the DQD transport in Fig. 1(b). This indicates that the donor site $\alpha$ is well separated from at least one of the drain and source leads. On the other hand, the signal of the donor site $\alpha$ strongly appears in the reflectometry spectrum [Fig. 1(c)], meaning the donor $\alpha$ has a large tunnel coupling to the source lead. By combining these insights, tunnel coupling between the donor $\alpha$ and the drain lead is determined to be suppressed. These geometrical characteristics are summarized into a hybrid DQD depicted in Fig. 2(b), where a negative $V_S$ drives electron transport through a sequential tunneling process, from source → donor $\alpha$ → QD → drain.

The bias triangle in Fig. 2(a) shows a weak peak feature along the base line of the triangle (filled red circle), which is more apparent in the spectrum measured with a high resolution at $V_S = -3$ mV [Fig. 3(a)]. This peak feature is attributed to resonant tunneling between the donor and the QD ground states. The bias triangle also shows transport structures attributed to the QD excited states. The open circle denotes the current onset due to the resonance between the donor ground state and a QD excited state. From its excitation energy of ~1 meV [equal to the source-drain bias in Fig. 2(a)] we can attribute this to a QD orbital excited state. The filled blue square, on the other hand, indicates a resonant tunneling structure involving another excited state. Interestingly, the excitation energy estimated at ~0.2 meV is much smaller than the orbital excitation energy just mentioned. If this would also be a QD orbital excited state, more excited states should be visible in the bias triangle, as the level spacings between orbital excited states do not change drastically in a QD formed by a smooth in-plane confinement well approximated by a parabolic potential. Therefore we conclude that the latter excited state belongs to the same orbital as the ground state, and its valley state is identified as the $|e\rangle$ state where the ground state has the $|g\rangle$ valley state. While Fig. 3(a) also shows a large transport signal in the black dotted box, this feature is the extension of an adjacent transport feature unrelated to the DQD of interest.

Fig. 3(e) shows the magnetic field dependence of the ground state peak as a function of back gate bias detuning $\Delta V_{BG}$ with respect to the valley excited state peak position at $V_{FG} = 76.5$ mV indicated by the vertical solid line. The ground state peak indicated by the vertical dashed line is suppressed with increasing magnetic field and disappears at ~200 mT. This magnetic field dependence indicates Pauli spin blockade (PSB) [28], which is consistent with the small intensity of the ground state peak. From the restriction in the electron configuration yielding PSB [Fig. 2(c)], the effective occupancies of the donor α and the QD, ($n_\alpha$, $n_{QD}$), are identified as (1, 1) [(0, 2)] before [after] the donor α → QD tunneling process.

The large intensity of the valley excited state peak provides an insight into the contribution of the valley degree of freedom to spin relaxation. PSB occurs because the exchange gap lifts the degeneracy of the spin singlet and triplet eigenstates in the ($n_\alpha$, $n_{QD}$) = (0, 2) configuration [23]. In a multi-valley QD system, the exchange gap is determined by the valley state in the (0, 2) charge configuration. The electron in the QD in the (1, 1) configuration is assumed to take the $|g\rangle$ valley state owing to the rapid valley relaxation. On this assumption, an electron tunneling to the QD $|e\rangle$ ($|g\rangle$) state results in the valley unpolarized (polarized) state $|ge\rangle$ ($|gg\rangle$) in the (0, 2) charge configuration, related to the excited (ground) state peak. While the valley polarized state keeps the spin singlet $|S_{gg}\rangle$ and the triplet $|T_{gg}\rangle$ eigenstates energetically gapped [Fig. 2(c)], the valley unpolarized state makes corresponding spin eigenstates $|S_{ge}\rangle$ and $|T_{ge}\rangle$ degenerate [Fig. 2(d)] [6]. As a consequence, transport through the QD $|e\rangle$ state is not blocked by the PSB mechanism, manifesting itself as a much larger peak than the ground state peak. We note that in the context of spin qubit readout based on PSB, a large valley splitting is desirable to suppress unwanted lifting of the blockade via the valley excited state.

We estimate the electric field dependence from a detailed analysis of the valley splitting. First we obtain precise peak positions by fitting a double peak function $g(V_{FG}, V_{BG}) = f_g(x_g) + f_e(x_e) + g_{offset}(V_{FG}, V_{BG})$, $x_i = V_{BG} - V_{BG,0,i} - c_i V_{FG}$ ($i = g, e$) to the current spectrum in the solid box in Fig. 3(a). For the single peak function $f_i(V_{FG}, V_{BG})$, we employ the convolution of a Lorentzian function $L(x) = Aw^2/(x^2 + w^2)$ and a triangle function $\Lambda(x) = (-x/l + 1)\theta(x)\theta(-x + l)$ [$\theta(x)$ is the Heaviside step function], because inelastic tunneling makes both peaks asymmetric [Fig. 3(b)]. $g_{offset}(V_{FG}, V_{BG})$ is an offset function to take account of single site transport through the donor site [29]. Figure 3(b) compares the data and $g(V_{FG}, V_{BG})$ as a function of $V_{FG}$ at $V_{BG} = 3.458$ and $3.472$ V, where the ground and excited state peaks are well developed, respectively. The obtained positions of the ground (excited) state peak are plotted as a dashed (solid) line in Fig. 3(a), while their fitting errors are shown by the grey-shaded area around these lines. These two lines have slightly different slopes, which indicates a change in valley splitting from one side of the triangle to the other. Figure 3(c) plots the obtained $V_{FG}$ dependence of the valley splitting. The lower horizontal axis shows the change in electric field perpendicular to the SOI interface $\Delta E_z$ estimated from $V_{FG}$ and $V_{BG}$ by a simple planar capacitor model [Fig. 3(d)]. In this model $\Delta E_z$ is represented as $\Delta E_z = (\Delta V_{BG} - \Delta V_{FG})/[t_{ch} + (t_F + t_B)\varepsilon_{Si}/\varepsilon_{SiO2}]$. Here we assume a channel thickness $t_{ch} = 20$ nm, front and back gate oxide thickness, $t_F = 5$ nm and $t_B = 145$ nm, and dielectric constants of Si and SiO$_2$, $\varepsilon_{Si} = 11.9$ and $\varepsilon_{SiO2} = 3.8$. The electric field dependence of the valley splitting is estimated at $1.2 \pm 0.2$ meV/(MV/m). This value is one order of magnitude larger than the previous report on a QD confined to a thermally defined Si/SiO$_2$ interface [0.13 and 0.27 meV/(MV/m) in Ref. 5] and comparable with that of a SOI structure defined by oxygen implantations [7]. The discrepancy from Ref. 5 can be explained by the profile of our back-side Si/SiO$_2$ interface, which can contain interface defects generated by hydrogen implantations in the wafer bonding process in comparison with the pure thermally defined Si/SiO$_2$ interface. The simplicity of our electrostatic model is also a possible explanation; it does not include the effect of the source and drain leads, possibly causing an underestimation of the electric field. Donors near the QD site may also enhance the electric field dependence of valley splitting via orbital mixing between the QD and donors [17, 21, 22].

In conclusion, we measured transport through a hybrid DQD consisting of a donor and a QD. The DQD transport spectrum reveals resonant tunneling features involving valley-split QD levels. The gate bias dependence of these features enables a direct measurement of the electric field dependence of the valley splitting in the QD. The electric field dependence of the valley splitting is estimated to be a larger value than previously reported [5]. We note that this discrepancy can be attributed to interface defects generated by hydrogen implantations, the underestimation of the electric field or orbital mixing with nearby donors.

**ACKNOWLEDGMENTS**

This work was supported by the ARC Centre of Excellence for Quantum Computation and Communication Technology (CE110001027), in part by the U.S. Army Research Office (W911NF-08-1-0527), and the European Community's seventh Framework under the Grant Agreement No. 318397 (TOLOP). S.R. acknowledges a Future Fellowship (FT100100589). M.Y.S. acknowledges a Laureate Fellowship. The device has been designed and fabricated by the AFSiD Project partners, see http://www.afsid.eu. The authors thank Joseph Salfi, and Benoit Voisin for fruitful discussions.

[29] To obtain good fitting, we use $g_{\text{offset}}(V_{FG}, V_{BG}) = [h(y) + b_h][\tanh(-x_g/w_b) + 1]/2 + b$, $h(y) = a_h[\tanh((y - y_l)/w_l) + 1][\tanh(-(y - y_r)/w_r) + 1]/4$, $y(V_{FG}, V_{BG}) = V_{FG} + c_h(V_{BG} - 3.48)$. $h(y)$ represents the broad structure lying almost vertically to the resonant tunneling peaks, attributed to the single site transport modulated by fluctuations of the density of state in leads. $a_h = 0.524$ pA, $y_l = 76.5$ mV, $y_r = 77.7$ mV, $w_l = 0.520$ mV, $w_r = 0.209$ mV, and $c_h = 10.0$ mV/V are obtained from the fitting to the structure in $3.478$ V $< V_{BG} < 3.496$ V, which is separately done from the fitting for the resonant peaks. To represent the fading of this structure around the base line, we multiply $(\tanh(-x_g/w_b) + 1)/2$. $a_h$, $b$ and $w_b$ are fitting parameters for the resonant peak fitting. The resulting $g_{\text{offset}}(V_{FG}, V_{BG})$ at $V_{BG} = 3.458$ V and $3.472$ V is plotted by dashed curves in Fig. 3(b).

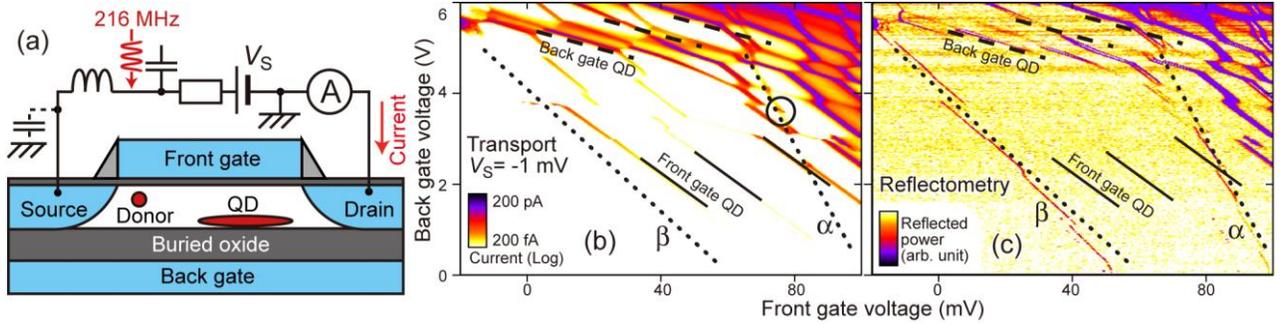

FIG. 1. (a) Schematic picture of the device structure and the measurement circuit. (b) Transport and (c) reflectometry spectra over a wide range of $V_{FG}$ and $V_{BG}$. Solid and dashed lines indicate features attributed to front gate side and back gate side QDs, respectively. Dotted lines show the positions where a reflectometry signal attributed to donor sites α and β appears in (c).

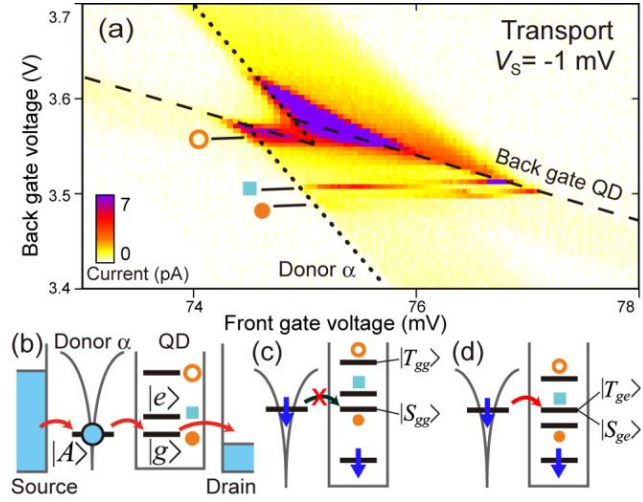

FIG. 2. (a) A bias triangle structure in the region enclosed by the circle in Fig. 1(b). Dashed and dotted lines, labeled back gate QD and α respectively, have the same slopes as those in Figs. 1(b) and (c). The filled circle, filled square and blank circle indicate three specific features in the bias triangle (see the text). (b) A schematic energy diagram of the hybrid DQD which provides the bias triangle in (a). The levels in the QD denoted by symbols involve the transport features indicated by the corresponding symbols in (a). (c) Spin-blocked tunneling related to the valley ground state. (d) Tunneling to the valley excited state. This tunneling is allowed because of the degenerate $|S_{ge}\rangle$ and $|T_{ge}\rangle$ states.

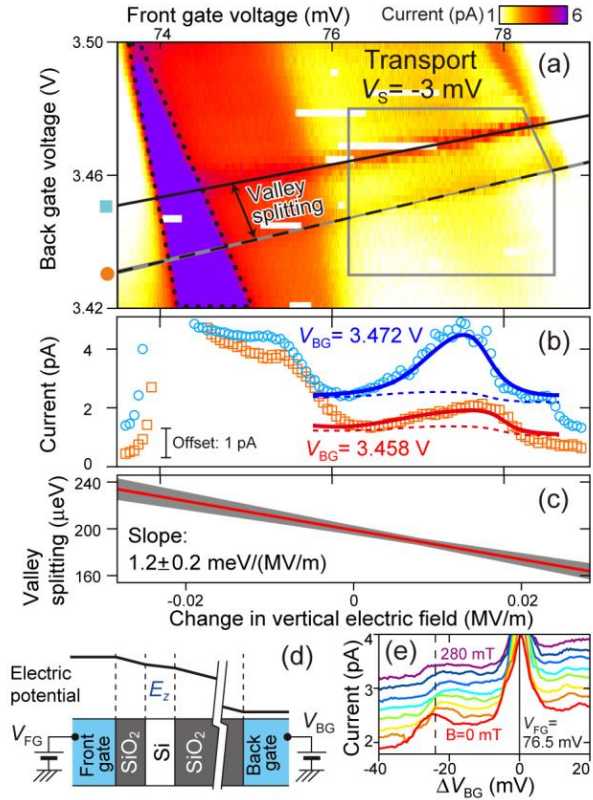

FIG. 3. (a) A high-resolution transport spectrum around the base line of the same bias triangle structure shown in Fig. 2(a), measured at $V_S$ = -3 mV. The positions of the peaks involving the valley excited (blue square) and ground (red circle) states are obtained from the fitting explained in the text and plotted as the solid and dashed lines, respectively. The grey-shaded area around the dashed line shows fitting error of the ground state peak position. The yellow area outlined by a dotted box indicates the structure attributed to the extension of an adjacent transport feature. (b) Slices of (a) at $V_{BG}$ = 3.458 V (square) and 3.472 V (circle). The plots are offset vertically by 1 pA for clarity. The solid curves show the fitting curves at these $V_{BG}$. $g_{\text{offset}}(V_{FG}, V_{BG})$ is plotted by the dashed curve at each $V_{BG}$. (c) The gate voltage dependence of the valley splitting. The grey-shaded area shows the fitting error. The bottom axis indicates the electric field at the corresponding gate voltage on the top axis. (d) The planar capacitor model used to estimate the electric field. The estimated change in the electric field is shown on the lower axis in (c). (e) Current measured at several magnetic field values (0 to 280 mT with 40 mT step) as a function of back gate bias detuning $\Delta V_{BG}$ with respect to the position of valley excited state peak (solid vertical line) at $V_{FG} = 76.5$ mV. The traces are offset by 0.2 pA vertically for clarity.